\documentclass[12pt,preprint,apj]{emulateapj}

\newcommand{\kms}{{km s$^{-1}$}}

\usepackage{epstopdf}
\usepackage{color}
\usepackage{amsmath}

\shorttitle{TRGB Distance to M60}
\shortauthors{Lee \& Jang 2017}

\begin{document}

\title{
Resolving the Discrepancy of Distance to M60, \\
a Giant Elliptical Galaxy in Virgo}
\author{Myung Gyoon Lee$^1$ and In Sung Jang$^{1,2}$}
\affil{$^1$Astronomy Program, Department of Physics and Astronomy, Seoul National University, Gwanak-gu, Seoul 151-742, Korea}
\affil{
$^2$Leibniz-Institut f\"{u}r Astrophysik Potsdam (AIP) An der Sternwarte 16, 14482 Potsdam, Germany}
\email{ mglee@astro.snu.ac.kr, isjang@aip.de}


\begin{abstract}
There is a well-known discrepancy in the distance estimation of M60, a giant elliptical galaxy in Virgo: the planetary nebula luminosity function (PNLF) distance moduli for this galaxy are, on average, $~$0.4 mag smaller than the values based on the surface brightness fluctuation (SBF) in the literature.
We present photometry of the resolved stars in an outer field of M60 based on deep F775W and F850LP images in the Hubble Space Telescope obtained as part of the Pure Parallel Program in the archive. Detected stars are mostly old red giants in the halo of M60. With this photometry, we determine a distance to M60 using the tip of the red giant branch (TRGB).
A TRGB is detected at F850LP$_{\rm TRGB}=26.70\pm0.06$ mag,
in the luminosity function of the red giants. This value corresponds to F814W$_{0,\rm TRGB}=27.13\pm0.06$ mag and $QT_{\rm RGB}=27.04\pm0.07$ mag, where $QT$  is a color-corrected F814W magnitude. From this we derive a distance modulus, $(m-M)_0=31.05\pm0.07{\rm(ran)}\pm0.06{\rm (sys)}$ ($d=16.23\pm0.50{\rm (ran)}\pm0.42{\rm (sys)}$ Mpc).
This value is 
 0.3 mag larger than the PNLF distances and 0.1 mag smaller than the SBF distances in the previous studies,  indicating that 
 the PNLF 
 distances to M60 in the literature have larger uncertainties than the suggested values. 
\end{abstract}

\keywords{galaxies: distances and redshifts --- galaxies: stellar content  --- stars : Population II --- galaxies: clusters: individual (Virgo, M60)  --- galaxies: elliptical and lenticular, cD } 

\section{INTRODUCTION}

Two of the popular distance indicators for nearby elliptical galaxies and early-type  spiral galaxies are the surface brightness fluctuation (SBF) \citep{ton88,ton01,bla09,bla10,can11,bla12} and
the planetary nebula luminosity function (PNLF) \citep{jac89,jac90,fel07,teo11,cia12,cia13}.  
The SBF method is based on the fact that the variance in the images of a galaxy depends on the distance to the galaxy.  
It can be applied to more distant galaxies compared with the tip of the red giant branch (TRGB) method, but its precision decreases for the galaxies with composite stellar populations \citep{bla12}.
The PNLF method is based on the estimation of the [OIII]$\lambda$5007 luminosity function of planetary nebulae (PNe) which is calibrated empirically. It is not supported by any theory, but it works well as a distance indicator. It is easy to apply, but does not work for galaxies with a small number of PNe \citep{cia12}.

There is a well-known discrepancy in the distance estimation based on these two methods in the sense that PNLF distance moduli are, on average, 0.3--0.4 mag smaller than SBF values
\citep{cia02,bla09,bla10,teo10,cia12,cia13}. 
\citet{cia12} and \citet{cia13} concluded in a review of the PNLF that this distance offset may be due to the combined result of several small effects including the difference in the zero points and the dust effect in spiral bulges  (see also \citet{can13}). The author pointed out that the main calibrators for the SBF and PNLF are intermediate-type spiral galaxies, while the targets are mostly early-type galaxies. 
It is also noted that the distance offset (between the SBF and PNLF distances) versus distance modulus diagram (see  Fig. 7 in \citet{cia12}) shows a trend that the scatter of this difference becomes larger at $(m-M)_0>31.0$.
To help resolve this discrepancy, distance estimation based on another independent method is needed for the common targets of the SBF and PNLF methods.

\begin{deluxetable*}{lcl}
\tablewidth{0pc} 
\tablecaption{Basic Parameters of M60\label{tab-info}}
\tablehead{
\colhead{Parameter} 	& \colhead{Value} & \colhead{References}}
\startdata
R.A.(J2000), Dec(J2000) 	& 12$^h$ 43$^m$ 40$^s$.0,  +11$^\circ$ 33$'$ 10$''$	& 1 \\
Morphological Type 					& E2 							& 2\\
Apparent total magnitude  	& $B^T = 10.24\pm0.03$ & 2\\
Apparent total color  	& $( B^T - V^T ) = 0.96\pm0.01$ & 2\\
Ellipticity	 			& 0.11 & 3\\
Position angle 			& 71 deg & 3\\
$D_{25}(B)$				&  $262\arcsec$ & 2\\ 
Effective radius		& $58\farcs7$ & 2 \\

Systemic velocity 		& 1110 km s$^{-1}$ & 1\\

Foreground extinction 	& $A_B=0.096$ $A_V=0.072$ $A_I=0.040$ & 5 \\
Distance modulus				& $(m-M)_0 = 31.05\pm0.07({\rm ran})\pm0.06({\rm sys})$ & 6 \\

Distance				& $d=16.23\pm0.50\pm0.42$ Mpc & 6 \\
Plate scale 				& 78.7 pc arcsec$^{-1}$ & 6\\
Absolute total magnitudes 		& $M_B^T = -20.92 $, $M_V^T = -21.85 $  & 2, 6 \\ 
Central velocity dispersion ($R_{\rm eff}/8$) & $213 $ km s$^{-1}$ & 4 \\
Dynamical mass for $R<8R_{\rm eff}$ &  $M=1.61\pm0.7 \times 10^{12} M_\odot$ & 7 \\ 
\enddata
\tablerefs{(1) NED; (2) \citet{dev91}; (3) \citet{mak14}; (4) \citet{cap13}; (5) \citet{sch11}; (6) This study; (7) \citet{ala16}.}
\label{tab_info}
\end{deluxetable*}

In this study we selected M60 (NGC 4649, VCC 1978), a giant elliptical galaxy in Virgo, to resolve the distance discrepancy between PNLF distances and SBF distances. Basic parameters of M60 are listed in {\bf Table \ref{tab-info}}.
M60 has been a target of numerous studies because it shows several interesting features.
First, it is the third brightest elliptical galaxy in Virgo, and hosts a rich population of globular clusters, PNe, and low-mass X-ray binaries (LMXBs) \citep{lee08a,lee08b,hwa08,str12,pot15,teo11,luo13,min14}.
Second, it has a small spiral companion, NGC 4647 (SAB(rs)c),
located at $2\farcm6$ (12 kpc) in the north-west from the center of M60. Whether M60 and NGC 4647 are interacting or not has been controversial \citep{deg06,lan13,dab14,min14,pot15}.
Third, it hosts a bright ultracompact dwarf (UCD), M60-UCD1, which is one of the densest galaxies \citep{str13,liu15}. A supermassive black hole (SMBH) with $M_{\rm SMBH}= 2.1\times 10^7 M_\odot$ was found in this UCD, and provides a strong evidence that this UCD is not a globular cluster, but a stripped nucleus of a genuine galaxy \citep{set14}.
Fourth, in the central region of M60  ($51\farcs6$
west and $78\farcs7$ south), an underluminous Type Ia supernova, SN2004W, was discovered \citep{moo04,nie12}. Unfortunately the full light curve of SN2004W is not available.
Fifth, it is a main member of  the M60 group, which includes M60, NGC 4647, M59 (NGC 4621), NGC 4660, and NGC 4638. This group may be  the nearest compact group of galaxies, if its nature as a genuine group is confirmed \citep{mam89, mam08}. 

In an extensive study of the PNe in M60, 
\citet{teo11} determined a distance to this galaxy using a large sample of PNe, and presented $(m-M)_0=30.7\pm0.2$ ($14.0\pm1.0$ Mpc).
This value is 0.4 mag smaller than the most recent value based on the SBF method, $(m-M)_0=31.1\pm0.2$ ($16.6\pm1.0$ Mpc) \citep{bla09}.  
M60 is an elliptical galaxy so that the dust effect for the PNLF and SBF distance estimation must be negligible. Therefore this discrepancy must be due to other effects, remaining to be explained.

In this study we analyse deep high resolution images of M60 available in the Hubble Space Telescope (HST) archive, and use them to determine a distance to M60 applying the TRGB 
\citep{lee93,riz07,jan17a}. 
%
Currently deriving TRGB distances to Virgo galaxies is difficult, but is possible with deep HST images.
To date, TRGB distances have been estimated only for a small number of Virgo galaxies:
M87 \citep{bir10,lee17} and 
several dwarf galaxies \citep{cal06,dur07,jan14}.
M60 is one of the rare examples in Virgo for which SBF, PNLF and TRGB distances can be compared.

This paper is organized as follows. In Section 2
we describe the reduction of data used in this study. Section 3 presents the color-magnitude diagrams (CMDs) of the resolved stars detected in the images of an outer field of M60. This is the first CMD of the resolved stars in M60.  Then we estimate a TRGB distance to M60 from photometry of these resolved stars. Section 4 compares  the TRGB distance determination results in this study and the PNLF and SBF distances in the previous studies, 
and discuss interaction between M60 and NGC 4647. In the final section, main results are summarized.

\begin{figure}
\centering
\includegraphics[scale=0.9]{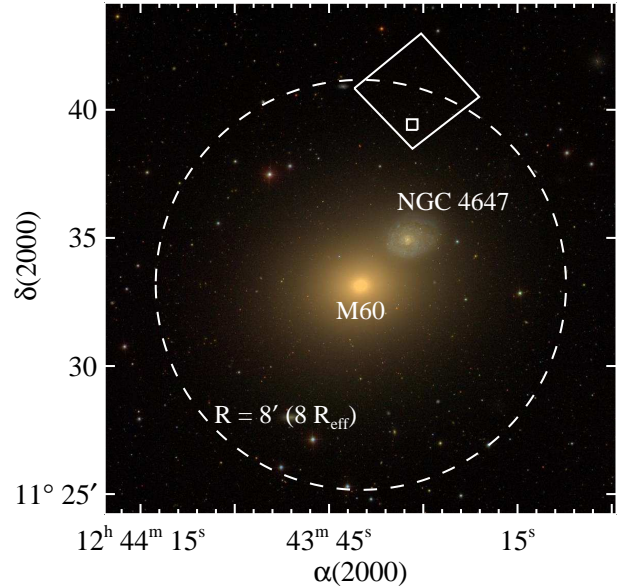}
\caption{Location of the HST field 
(a large box) 
at $R\approx 8'$ in the north from the center of M60 on the color image in the SDSS. North is up, and east to the left. 
 A small square  
in the HST field indicates the location of a $10\arcsec \times 10\arcsec$ gray scale map shown in Figure \ref{fig_map}.
A spiral galaxy in the north-west of M60 is NGC 4647.
Whether NGC 4647 is interacting with M60 has been controversial.
}
\label{fig_finder1}
\end{figure}

\begin{figure}
\centering
\includegraphics[scale=0.9]{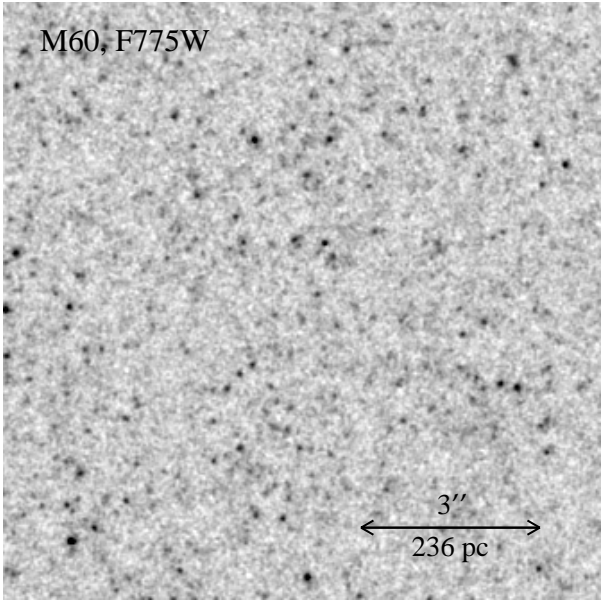} 
\caption{A grayscale map of F775W image of a $10'' \times 10''$ field for M60, which is marked by a small square in Figure 1. Most of the point sources in this map are old red giant stars belonging to M60.
}
\label{fig_map}
\end{figure}

\section{DATA and DATA REDUCTION}

{\bf Figure \ref{fig_finder1}} displays a finding chart for a $20' \times 20'$ field including M60 based on the SDSS color map. It shows also NGC 4647, a spiral galaxy in the northwest of M60.  
We used the ACS/WFC F775W (SDSS $i'$) and F850LP (SDSS $z'$) images of an outer field in the north of M60, 
the location of  which  is marked in  {\bf Figure \ref{fig_finder1}}.  
These images were obtained as part of the ACS Pure Parallel Program (PID:9575, PI. William Sparks) in 2002. However, they are very useful for the study of the resolved stars in M60 as well.

Since the difference between the effective wavelengths of F775W and F850LP filters is small, the combination of these two filters is not effective for the study of colors (or metallicity) of stellar populations. However this combination of the filters is good enough to detect red giants in nearby galaxies. 

The HST field is located at $\approx 8'$ in the north from M60 in the sky.
The effective radius of M60 is $R_{\rm eff} = 58\farcs7$ \citep{dev91} so the projected galactocentric distance of the HST field is
about 8 $R_{\rm eff}$.  Therefore the crowding of the point sources in this field is much lower compared with inner fields so that this field is much more suitable for the study of resolved stars in M60. 
We combined individual exposure images to produce deep master images using the AstroDrizzle package.
Total exposure times are 15,407 s for F775W
and 9,547 s for F850LP so that the images are deep enough to study the resolved stars in M60.
A gray scale map of the F775W image for a $10'' \times 10''$ section of the entire field (marked by a small square in {\bf Figure 1})  is shown in {\bf Figure \ref{fig_map}}. In the figure, many point sources are clearly seen. Most of them are red giant stars belonging to M60, and some of them may be compact background galaxies.

\begin{figure}
\centering
\includegraphics[scale=0.9]{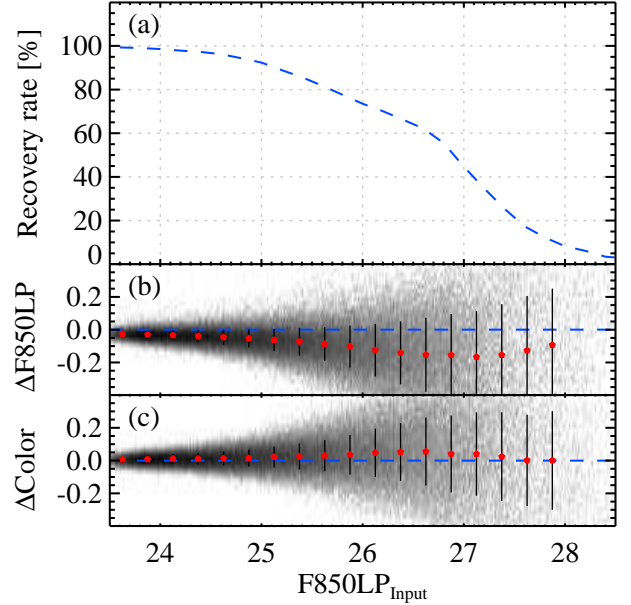} 
\caption{
 (a) Recovery rates of red giant branch stars with $0.3 \leq$ (F775W--F850LP) $\leq 0.9$, derived from artificial star experiments. (b) Differences between input and output F850LP magnitudes (input minus output) as a function of input F850LP magnitudes. Median offsets and standard deviations in each magnitude bin are indicated by red dots and vertical lines, respectively. (c) Same as (b) except for (F775W--F850LP) colors. 
}
\label{fig_ast}
\end{figure}

 We obtained photometry of the point sources in the images using the latest version of DOLPHOT \citep{dol00}.
We used charge transfer efficiency corrected and flat-fielded images (*\_flc.fits images) with the synthetic Tiny Tim point spread functions (PSFs) \citep{kri11}. The DOLPHOT parameters used in this study are the same as those given in DOLPHOT/ACS user's guide (version 2.0). 

 We carried out artificial star tests using the artificial star routine (acsfakelist) in DOLPHOT. 
We generated a sample of artificial stars having a color range of F775W--
F850LP = 0.3 $\sim$ 0.9 mag and a magnitude range of F850LP = 23.0 $\sim$ 29.0 mag. We added 10,000 artificial stars, which corresponds to $\sim$ 10\% of the total number of detected sources, in each image, and carried out PSF photometry as done on the original frames. We iterated this procedure 50 times
to reduce statistical uncertainties. 
{\bf Figure \ref{fig_ast}} displays the recovery rates of the input stars, and the difference in F850LP magnitudes and (F775W-F850LP) colors between the input and output values. It shows that 50\% completeness limit is F850LP $\sim$26.9 magnitude. The mean values of the input minus output magnitudes and colors are 
$\Delta$F850LP $= -0.158\pm0.007$ mag and $\Delta$(F775W--F850LP) $= 0.057 \pm0.007$ mag, for F850LP = 26.8 mag which is close to the TRGB magnitude.


\begin{figure}
\centering
\includegraphics[scale=0.9]{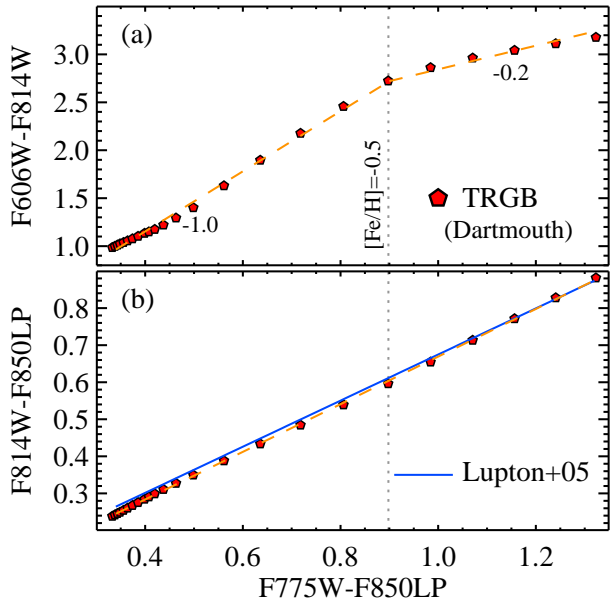} 
\caption{
(a) (F606W--F814W) vs. (F775W--F850LP)  relation. 
Pentagons denote the TRGB color from the 12 Gyr Dartmouth model with $-2.3\leq$[Fe/H]$\leq 0.0$
(the data for [Fe/H] = --1.0, --0.5, and --0.2 are labeled).
Dashed lines represent the linear fits.
(b) Same as (a) but for (F814W--F850LP) vs. (F775W--F850LP)  relation.
The blue solid line represents the relation between Jonhson/Cousins and SDSS system given by \citet{lup05}.
}
\label{fig_transform}
\end{figure}

Since the TRGB calibration is based on $VI$ (or F606W and F814W in the HST system) \citep{lee93,riz07,jan17a}, we need to transform F775W and F850LP photometry to 
F606W and F814W photometry.
For this purpose we used the 12 Gyr isochrones in the Dartmouth model \citep{dot08}.
In {\bf Figure \ref{fig_transform}} we plotted the color-color  relations for the TRGB of the isochrones for a range of metallicity ($-2.3\leq$[Fe/H]$\leq 0.0$): 
(a) (F606W--F814W) versus (F775W--F850LP)  relation, and 
(b) (F814W--F850LP) versus (F775W--F850LP)  relation. 
The (F606W--F814W) versus (F775W--F850LP)  relation is fit well by a double linear relation with a break at (F775W--F850LP)$=0.9$ (corresponding to [Fe/H] =  --0.5), while 
the (F814W--F850LP) vs. (F775W--F850LP)  relation is represented well by a single linear relation 
 for (F775W--F850LP)$<1.35$.
From the linear fits for the data, we obtain

\begin{equation}
\begin{aligned}
(F606W-F814W) = (3.136\pm0.014)(F775W-F850LP) \\
-(0.102\pm0.007) 
\end{aligned}
\end{equation}
 with $rms=0.034$ for  (F775W--F850LP)$\le0.9$,  and 
 \begin{equation}
 \begin{aligned}
(F606W-F814W) = (1.235\pm0.021)(F775W-F850LP) \\
+(1.610\pm0.019)
\end{aligned}
\end{equation}
  with $rms=0.029$ for  (F775W--F850LP)$>0.9$.
  
Similarly, we derive 
 \begin{equation}
 \begin{aligned}
(F814W- 850LP) = (0.642\pm0.006)(F775W-F850LP) \\
+(0.027\pm0.004) 
\end{aligned}
\end{equation}
 with $rms=0.003$.

On the other hand, \citet{lup05} derived, from the comparison of SDSS photometry and Johnson-Cousins photometry of standard stars,  a transformation relation between the two systems: 
$I = i - 0.3780\times(i - z)  -0.3974$ (rms = 0.0063), as plotted by the blue solid line in {\bf Figure \ref{fig_transform}(b)}.
The second relation derived in this study is very similar to this transformation,  
except for the slight offset in the blue end. 
 Using the equations above, we can transform (F775W--F850LP) colors and
F850LP magnitudes of the detected stars in the HST field of M60 to  (F606W--F814W) colors and F814W magnitudes.

\section{RESULTS}

\begin{figure}
\centering
\includegraphics[scale=0.9]{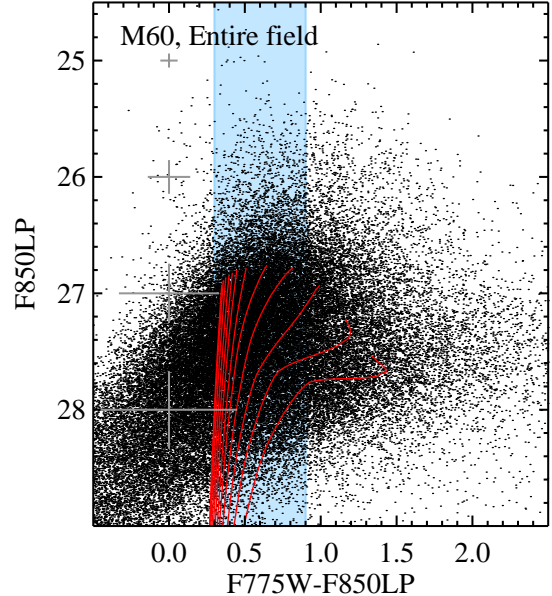} 
\caption{F850LP--(F775W-F850LP) color-magnitude diagram for the resolved point sources in the HST field of M60. 
Curved lines represent 12 Gyr stellar isochrones with a range of metallicity ([Fe/H] =--2.2 to 0.0, in steps of 0.2, from left to right) in the Dartmouth models \citep{dot08}. The vertical shaded region at 0.3 $\leq$ (F775W--F850LP) $\leq$ 0.9 indicates the region for the blue RGB stars, which we used in the TRGB analysis.  
The errorbars at (F775W-F850LP)=0.0 represent the mean errors of F850LP magnitudes and (F775W-F850LP) colors of the detected stars in the shaded region. 
}
\label{fig_cmd}
\end{figure}

\subsection{CMDs 
of the Resolved Stars in M60}

In {\bf Figure \ref{fig_cmd}} we plotted the F850LP--(F775W--F850LP) CMD of the detected point sources in the HST field of M60.
The most prominent feature in the CMD is a concentration of red stars with a broad range of colors, the mean value of which is (F775W--F850LP) $\approx 0.6$. 
It is a red giant branch (RGB) of M60. 
The brightest part of this RGB is seen at F850LP$\approx 26.8$ mag, 
which corresponds to the TRGB of M60.  
The number density of the stars above the TRGB is much lower than that below the TRGB.
Our photometry of the resolved stars goes more than one magnitude below the TRGB so it can be used for reliable TRGB distance estimation of M60.

The width of the bright RGB with F850LP$<27.0$ mag 
is much larger than the mean photometric errors of the colors so it is mainly due to a large range of metallicity of the RGB stars in M60.
 We overlayed 12 Gyr stellar isochrones with a range of metallicity ([Fe/H] = --2.2 to 0.0, in steps of 0.2) in the Dartmouth models \citep{dot08}, shifted according to the distance to M60, by red lines. It is seen that the broad RGB of M60 is roughly overlapped by the RGB part of the isochrones with a range of metallicity.

\subsection{TRGB Distance Estimation}

\begin{figure}
\centering
\includegraphics[scale=0.9]{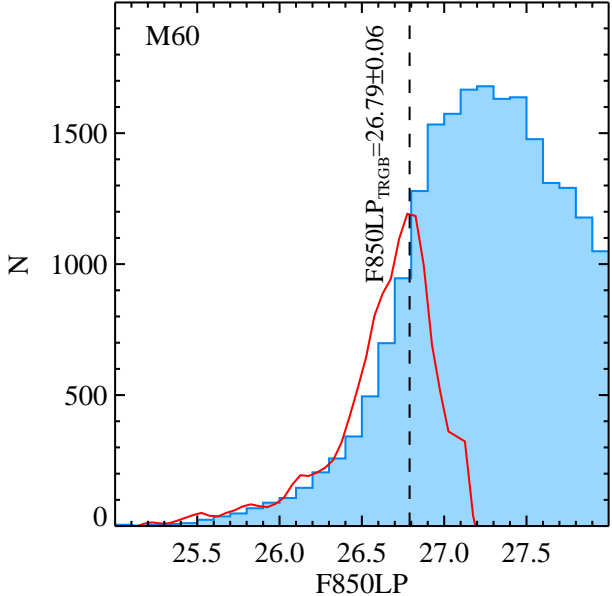}
\caption{F850LP luminosity function for blue RGB stars in M60 (histogram) and corresponding  edge detection response (red line). 
The position of the TRGB is marked by the dashed line.
}
\label{fig_lf}
\end{figure}

We determine a TRGB distance to M60 from photometry of the resolved stars, as done in our previous studies for other galaxies \citep{lee16, jan17b}.
 {\bf Table \ref{tab_distance}} lists a summary of TRGB distance estimation for M60.
First, we selected the blue red giant candidates inside the shaded region in the CMD of {\bf Figure \ref{fig_cmd}}. The TRGB magnitude is almost constant in this blue RGB. The data for M60 used in this study are not deep enough to cover the full range of colors of the RGB stars. In this case, using the blue RGB is the best way to avoid any complications due to redder stars. 
Then we derived their luminosity functions as shown in {\bf Figure \ref{fig_lf}}. Applying the edge-detection method with a Sobel kernel [--1, --2, --1, 0, 1, 2, 1] 
 to this luminosity function, we calculated the edge-detection responses, as plotted by the red solid line in the figure.
The edge-detection response shows clearly a major single peak at F850LP $\sim$26.8 mag.

 A quantitative value for the TRGB magnitude and its error were estimated using the bootstrap resampling method, as done in \citet{jan17b}. We performed ten thousand simulations of bootstrap resampling. In each simulation, we resampled a half number of stars randomly from the original sample and measured the TRGB as done for the original sample. Then we performed a Gaussian fit to the measured TRGB magnitudes and quoted the Gaussian mean for the mean TRGB magnitude and the width for the TRGB measurement error. 
This process gives a TRGB magnitude of F850LP$_{TRGB}$ = $26.79\pm0.06$ mag.
The median color of the TRGB, (F775W--F850LP)$_{TRGB}$ = $0.63\pm0.02$,  is measured using the RGB stars at $\pm$0.02 mag range of the TRGB. 
We estimated the systematic offsets of the TRGB magnitude and color using the artificial stars that have a luminosity function with a logarithmic slope of $\alpha =0.3$ and 
F850LP$_{TRGB} = 26.70$ mag and
(F775W--F850LP)$_{TRGB}$ = $0.60$.
We derived the TRGB magnitude and color from the recovered artificial stars using the same procedure. 
The mean values of the input minus output TRGB magnitudes and colors are 
$\Delta$F850LP$_{TRGB}= -0.09\pm0.06$ mag and $\Delta$(F775W--F850LP)$_{TRGB}= 0.06 \pm0.02$ mag.
Correcting the measured TRGB values with these systematic offsets, 
we obtain 
F850LP$_{TRGB}$ = $26.70\pm0.06$ mag and (F775W--F850LP)$_{TRGB}$ = $0.69\pm0.02$.

\begin{deluxetable}{lcc}
\tabletypesize{\footnotesize} 
\setlength{\tabcolsep}{0.05in}
\tablecaption{A Summary of TRGB Distance Measurement for M60}
\tablewidth{0pt}

\tablehead{ \colhead{Parameter}  &  \colhead{Value} 
} 
\startdata

Apparent TRGB magnitude in F850LP				&  $26.79\pm0.06$ \\
Apparent TRGB color in F775W--F850LP			&  $0.63\pm0.02$ \\
Systematic offset in F850LP						& $-0.09\pm0.06$ \\
Systematic offset in F775W--F850LP				&  $0.06\pm0.02$ \\
Corrected TRGB magnitude in F850LP				&  $26.70\pm0.06$ \\
Corrected TRGB color in F775W--F850LP			&  $0.69\pm0.02$ \\
Foreground extinction at F775W	& 0.043 \\
Foreground extinction at F850LP	& 0.033 \\
Intrinsic TRGB magnitude in	F850LP		& $26.67\pm0.06$ \\
Intrinsic TRGB color in	F775W--F850LP	& $0.68\pm0.02$ \\

Intrinsic TRGB magnitude in F814W					& $27.13\pm0.06$\\
Intrinsic TRGB color in F606W--F814W				& $2.03\pm0.06$\\
Intrinsic TRGB magnitude in $QT$					& $27.04\pm0.07$\\

Absolute TRGB magnitude		&$-4.015\pm0.057$ \\
Distance modulus, $(m-M)_0$					& $31.05\pm0.07_r\pm0.06_s$ \\
Distance, $d$ [Mpc]					&  $16.23\pm0.50_r\pm0.42_s$\\

\enddata
\label{tab_distance}
\end{deluxetable}

Then we corrected these for the foreground extinction effect,
using the values in \citet{sch11}, $A_{\rm F775W} = 0.043$ and $A_{\rm F850LP} = 0.033$.
M60 is a typical elliptical galaxy and our HST field is far from the M60 center so that it is expected that our HST field contains little dust. Indeed no far-infrared emission is detected in M60 \citep{lan13}. Thus we ignore internal reddening for M60 in this analysis.
We converted the measured TRGB magnitude and color in the F775W and F850LP system to the F606W and F814W system using the photometric transformations described in Section 2, obtaining  F814W$_{0,TRGB}=27.13\pm0.06$ mag
 and (F606W--F814W)$_{0,TRGB} = 2.03\pm0.06$. 

It is known that the $I$-band TRGB has a weak metallicity dependence, especially at the red color range (F606W--F814W $\gtrsim$ 1.5) \citep{bel01, riz07, jan17a}. 
\citet{jan17a} introduced a color-dependence-corrected TRGB magnitude, called as the $QT$ magnitude.
It is described by $QT$ = F814W$_0 - 0.159(Color-1.1)^2 + 0.047(Color - 1.1)$, where $Color$ = (F606W--F814W)$_0$. The absolute zero-point of the $QT$ is measured to be $M_{QT,TRGB}=-4.015\pm0.056$ mag, from the combination of two distance anchors with known geometric distances (NGC 4258 and the LMC). The systematic error of $\pm0.056$ in this calibration is much smaller than the values given in the previous studies. 
The value of the $QT$ magnitude and corresponding distance modulus for M60 we obtained are: $QT_{RGB} = 27.04\pm0.07$ mag and $(m-M)_0=31.05\pm0.07$ (random) mag (d = $16.23\pm0.50$ Mpc). 
The systematic uncertainty of this distance modulus is $\pm0.056$ mag (corresponding to the distance error of $\pm0.42$ Mpc).

\section{DISCUSSION}

\subsection{Comparison of the TRGB distance to M60 with PNLF and SBF Distances}

\begin{deluxetable*}{llll}
\tablewidth{0pc} 
\tablecaption{Comparison of Distance Estimates for M60\label{tab-comp}}
\tablehead{
\colhead{Method} 	& \colhead{$(m-M)_0$} & \colhead{References} & \colhead{Remarks}}
\startdata
TRGB			& $31.05\pm0.07({\rm ran})\pm0.06({\rm sys})$ & This study \\
\hline 
PNLF				& $30.76\pm0.08({\rm ran})\pm0.13({\rm sys})$ & \citet{jac90} &  N(PN)=16, $M^*_{\rm PN} = -4.48$  \\
 				& $30.73^{+0.10}_{-0.13}$ & \citet{cia02} & N(PN)=16, $M^*_{\rm PN} = -4.48$\\ 
				& $30.7\pm0.2$ 
				& \citet{teo11} &  N(PN)=218, $M^*_{\rm PN} = -4.48$ \\
									& $30.8\pm0.2$ 
				& \citet{cia13} &  N(PN)=218, $M^*_{\rm PN} = -4.54\pm0.05$   \\
  &  $30.74\pm0.09$ &  Mean &   \\
				\hline
				
SBF			& $31.06\pm0.11$ & \citet{nei00} &  F814W \\ 
			& $31.13\pm0.15$ & \citet{ton01} & $I$\\
			& $31.19\pm0.07({\rm ran})\pm0.15({\rm sys})$ & \citet{mei07} & F850LP \\
			& $31.08\pm0.08({\rm ran})\pm0.15({\rm sys})$ & \citet{bla09} &  F850LP\\
			  &  $31.13\pm0.05$ & Mean &   
\enddata
\label{tab_comp}
\end{deluxetable*}


\begin{figure}
\centering
\includegraphics[scale=0.9]{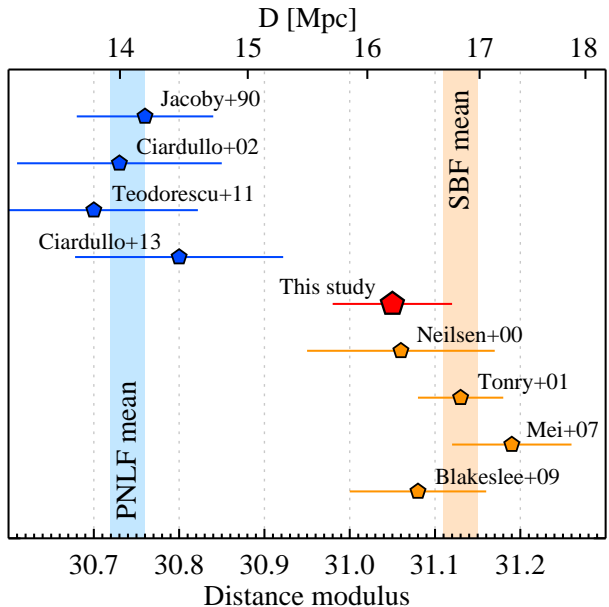} 
\caption{Comparison of the TRGB distance (red pentagon) in this study with the PNLF (blue pentagons) and SBF distances (yellow pentagons) in the literature. Error bars indicate standard errors. Mean values of the PNLF and SBF distances are marked by the blue and yellow vertical strips, respectively.
}
\label{fig_distcomp}
\end{figure}

We compared our TRGB distance estimate for M60 with those based on the PNLF and SBF in the literature, as summarized 
in {\bf Table \ref{tab_comp}} and plotted in {\bf Figure \ref{fig_distcomp}}.
\citet{jac90} presented a PNLF  distance to M60 derived from a small sample of 16 PNe, $(m-M)_0=30.76\pm0.14$ 
($d=14.2\pm0.6$ Mpc) including a systematic error of 0.13, which was updated later to 
$(m-M)_0=30.73^{+0.10}_{-0.13}$ by \citet{cia02}.
They adopted the calibration for the PNLF given by \citet{cia89}, $M^*_{\rm PN} = -4.48$, which is based on the Cepheid distance to M31, 710 kpc ($(m-M)_0=24.26\pm0.10$) and foreground reddening $E(B-V)=0.11\pm0.02$.
The distance to M31 adopted for this PNLF  calibration is somewhat smaller than the values in more recent Cepheid distances to M31: $(m-M)_0=24.51\pm0.08$ and $(m-M)_0=24.32\pm0.09$ (e.g, in \citet{wag15}). 

Later \citet{teo11} used a much larger sample of 218 PNe in M60, obtaining a similar value, $(m-M)_0=30.70\pm0.20$, including a systematic error of 0.13. They adopted the same calibration for the PNLF as used in \citet{jac90}. 
If the metallicity dependence of the period-luminosity relation for Cepheids is adopted for bright galaxies, this calibration will be slightly brighter to $M*_{\rm PN} = -4.53\pm0.04$ \citep{cia13}.  \citet{cia13} presented a similar calibration for metal-rich galaxies (with higher metallicity than that of the LMC), adopting the TRGB distances in \citet{tul09}: $M^*_{\rm PN} = -4.54\pm0.05$.
If this brighter calibration is used, the PNLF distance modulus for M60 will increase by 0.05--0.06 mag.
The mean value of these PNLF distances is derived to be $(m-M)_0=30.74\pm0.09$.

On the other hand, \citet{nei00} presented an SBF distance to M60, $(m-M)_0=31.06\pm0.11$. This value is consistent with the value given by \citet{ton01}, $(m-M)_0=31.13\pm0.05$.
Later \citet{mei07} presented a similar value, $(m-M)_0=31.19\pm0.07$.
\citet{bla09} updated the distances in \citet{mei07} with an improved calibration, presenting a 0.11 mag smaller value for M60, $(m-M)_0=31.08\pm0.08$. 
This is the most recent value for the SBF distance to M60 available in the literature.
The mean value of these SBF distances is derived to be $(m-M)_0=31.13\pm0.05$.
Our TRGB distance modulus, $(m-M)_0 = 31.05\pm0.09$,
is $\sim$0.3 mag larger 
than the mean PNLF value, and $\sim$0.1 mag smaller 
than the mean SBF value. 
This indicates that the PNLF distances to M60 in the literature have larger uncertainties than the suggested values.

\citet{cia13} pointed out that two main causes for the discrepancy between the PNLF and SBF distances are zero-points in the calibration and
the extinction effect due to dust in spiral galaxies. In the case of M60, the dust extinction is negligible. Then only the calibration problem remains for M60.
Note that the galaxies used for the calibration of the PNLF and SBF are mostly late-type galaxies, while a significant fraction of the galaxies used for the comparison of the PNLF and SBF distances are early-type galaxies \citep{cia13}. 
It is needed in the future to check any possible difference in the calibration of the PNLF and SBF method between the late-type galaxies and early-type galaxies. 

The result in this study is based on only one galaxy so that it may be too early to resolve the discrepancy between the PNLF distances and SBF distances. However, the result for M60 in this study will serve as a precious data point to understand the causes for the discrepancy.

\subsection{Interaction between M60 and NGC 4647}

NGC 4647, a spiral galaxy, is located only  $2\farcm6$ (corresponding to a projected distance of 12 kpc) in the north-west from the center of M60 in the sky. This pair of galaxies is called, Arp 116, and is a rare example of a combination of an elliptical galaxy and a spiral galaxy. The heliocentric radial velocity of NGC 4647 ($1409\pm1$ \kms, NED) is only about 300 \kms~ larger than that of M60 ($1110\pm5$ \kms, NED).
Because of the projected proximity  and the small radial velocity difference of M60 and NGC 4647, several studies investigated any possibility of tidal interaction between these two galaxies \citep{deg06,lan13,dab14,min14,pot15}.
However whether NGC 4647 is interacting with M60 or not is still controversial (\citet{deg06}, \citet{pot15} and references therein).

Optical images of this pair of galaxies, 
in {\bf Figure \ref{fig_finder1}}, show little evidence for any significantly distorted structures around each galaxy, although they are close to each other in the sky. This indicates two possibilities. First, the relative distance along the line of sight in the space between the two galaxies is so large that they are not interacting.
Second, they are relatively close to each other in the space, but their interaction is weak.
Recently \citet{pot15} found, from the study of kinematics of the globular clusters in M60, no strong evidence to support the interaction between M60 and NGC 4647. 
They suggested that M60 and NGC 4647 may be only in the beginning stage of interaction, if they are interacting, as suggested earlier by \citet{deg06} who noted a presence of weak young blue stellar population in the north-west direction of M60 in the HST/ACS images of the central region.

 It is known that strongly-interacting galaxies show relatively stronger MIR and FIR emission than weakly-interacting galaxies so the spectral energy distributions (SEDs) of galaxies are useful to estimate the stage of interaction \citep{lan13}. 
\citet{dop02} presented a five-stage classification scheme to estimate the stage of galaxy interaction. According to this scheme, weakly-interacting galaxies (Stage 2) show minimal morphological distortions, and
moderately-interacting galaxies (Stage 3) show stronger morphological distortion including often tidal tails \citep{lan13}. 
\citet{lan13} suggested, from the SEDs of NGC 4647 and M60 based on UV to FIR data, that this pair of galaxies is in the moderately interacting stage.

We need to know a relative distance between M60 and NGC 4647 to conclude on the possibility of tidal interaction between the two galaxies. 
Unfortunately there is not yet any TRGB distance  to NGC 4647.
There are several estimates for the distance to NGC 4647 based on the Tully-Fisher relation in the literature, showing a large range: $(m-M)_0 = 31.02\pm0.54$ to $31.95\pm0.30$ (see the compilation of the distance estimates in NED).
\citet{sol02} derived a mean value of the previous estimates, $(m-M)_0 = 31.25\pm0.26$ ($d=17.78\pm2.13$) (see the references in \citet{sol02}).
Recently \citet{tul13} presented an updated value based on the Tully-Fisher relation, $(m-M)_0 = 31.06\pm0.20$ ($d=16.3$ Mpc), which is 0.2 mag smaller than the value in \citet{sol02}. 
This value for NGC 4647 is only 0.01 mag larger than the TRGB distance to M60, indicating that 
both galaxies may be located at the similar distance.
However, absence of any significantly distorted structures around each galaxy indicates that the relative distance between the two is not close enough to show strong tidal interaction.
The error in the current Tully-Fisher distance to NGC 4647 is as large as $\pm0.2$ mag ($\pm1.5$ Mpc in the linear scale).
A more precise estimation of the distance to NGC 4647 is needed to draw a solid conclusion on the tidal interaction between M60 and NGC 4647.

\section{SUMMARY}

We present photometry of the resolved stars in an outer field of M60 based on deep F775W and F850LP images. This is the first photometry of the resolved giant stars in M60. 
Primary results in this study are summarized as follows.

\begin{enumerate}

\item The CMD of the resolved stars in M60 shows a distinguishable broad RGB. 
A TRGB is detected at F850LP$_{\rm TRGB}=26.70\pm0.06$ mag, 
in the luminosity function of the red giants. 
This value corresponds to F814W$_{\rm 0,TRGB}=27.13\pm0.06$ mag and $QT_{\rm RGB}=27.04\pm0.07$ mag.

\item From the magnitude of the TRGB we derive a distance modulus, $(m-M)_0=31.05\pm0.07{\rm(ran)}\pm0.06{\rm (sys)}$ (the total error is $\pm0.09$). 
The corresponding linear distance is  $d=16.23\pm0.50{\rm(ran)}\pm0.42{\rm (sys)}$ Mpc
(the total error is $\pm0.65$ Mpc).

\item The TRGB distance modulus for M60 derived in this study is 0.3 mag larger than the mean PNLF distance values, and 0.1 mag smaller than the SBF distance values.
This indicates that 
the PNLF 
distances to M60 in the literature have larger uncertainties than the suggested values.




\item We checked the relative distance between M60 and NGC 4647, a nearby spiral galaxy, to investigate any tidal interaction between the two galaxies.
It is found that the TRGB distance to M60 and the Tully-Fisher distance to NGC 4647  are similar within the errors.
However, absence of any significantly distorted structures around each galaxy indicates that the relative distance between the two is not close enough to show strong tidal interaction.

\end{enumerate}

\bigskip
This study is supported by the National Research Foundation of Korea (NRF) grant
funded by the Korea Government (MSIP) (No. 2012R1A4A1028713). 
This paper is based on image data obtained from the Multimission Archive at the Space Telescope Science Institute (MAST).

%
%
%

\end{document}